\documentclass[aps,prl,reprint,groupedaddress,floatfix,showpacs]{revtex4-1}
\usepackage{amsmath}
\usepackage{graphicx}
\usepackage{bm}
\usepackage[caption=false]{subfig}
\usepackage{microtype}

\begin{document}

\title{Turbulent Steady States in Two-Dimensional Sonic Black Holes: Superfluid Vortices and Emission of Sound}

\author{R. B\"urkle}
\email{rbuerkle@physik.uni-kl.de}
\author{A. Gaidoukov}
\author{J. R. Anglin}
\affiliation{Fachbereich Physik, Technische Universit\"at Kaiserslautern, D-67663 Kaiserslautern, Germany}

\date{\today}

\begin{abstract}
Simulation of a sonic black-hole/white-hole pair in a (2+1)-dimensional Bose-Einstein condensate shows formation of superfluid vortices through dynamical instabilities seeded by initial quantum noise. The instabilities saturate in a quasi-steady state of superfluid turbulence within the supersonic region, from which sound waves are emitted in qualitative resemblance to Hawking radiance. The power spectrum of the radiation from the slowly decaying two-dimensional sonic black hole is strongly non-thermal, however.
\end{abstract}


\maketitle
A black hole is among the simplest possible steady states in pure general relativity. Hawking's seminal 1974 paper \cite{Hawking} showed, however, that adding quantum fluctuations changes the black hole steady state into a thermally radiating one. The further addition of gravitational backreaction is expected to make this state only quasi-steady, as the radiating black hole slowly (at least initially) shrinks. Hawking's revised picture of the black hole steady state has suggested a connection between gravity, quantum mechanics, and thermodynamics, but the result is uncertain because new physics at trans-Planckian frequencies might radically revise the conclusion. 

To gain insight into this potential sensitivity to short-distance physics, Unruh proposed studying sonic horizons in fluids \cite{Unruh}, where the trans-Planckian onset of quantum gravity has the analogy of hydrodynamics breaking down at short distances. The \textit{ergoregion} inside a black hole corresponds, in this analogy, to a region of locally supersonic fluid flow; the surface at which the flow passes through the local speed of sound is the sonic horizon. The physics of sonic horizons has since grown into a topic in its own right \cite{Volovik_book,Unruh_sonic, Unruh2007,unruh_water2, water}, and it has recently been claimed that analog Hawking radiation has been observed \cite{Steinhauer_1, Steinhauer_2, Steinhauer_3} in a quasi-one-dimensional Bose-Einstein condensate (BEC).

In this Letter we simulate the time evolution of an initial horizon configuration in a spatially two-dimensional BEC, incorporating nonlinear back-reaction at the classical level (\textit{i.e.}, in Gross-Pitaevskii mean field theory) while modelling quantum vacuum fluctuations with Gaussian noise in the initial state. We find that linear dynamical instability of the initial laminar flow, as identified within Bogoliubov-de Gennes perturbation theory for BEC backgrounds with finite ergoregions \cite{Garay, Garay2}, leads in two spatial dimensions to the formation of quantized vortices. After an initial transient epoch, the system relaxes to subsonic flow through a long quasi-stationary phase in which vortices and non-thermal sonic noise are steadily emitted from the turbulent ergoregion.

This non-thermal turbulence radiation constitutes a counterexample to the hypothesis that strictly thermal Hawking radiation at long wavelengths is insensitive to short-distance physics. At the same time we set Hawking radiation itself within a larger context of non-trivial steady states of horizons. Even if sonic black holes turn out to be qualitatively different from real ones, the relationships between geometry and equilibration may become clearer within this wider frame of horizon physics.

\paragraph{\label{setup} Setup.---}
\begin{figure}
\centering
\includegraphics[width=0.45\textwidth, trim=0mm 0mm 0mm 0mm ,clip]{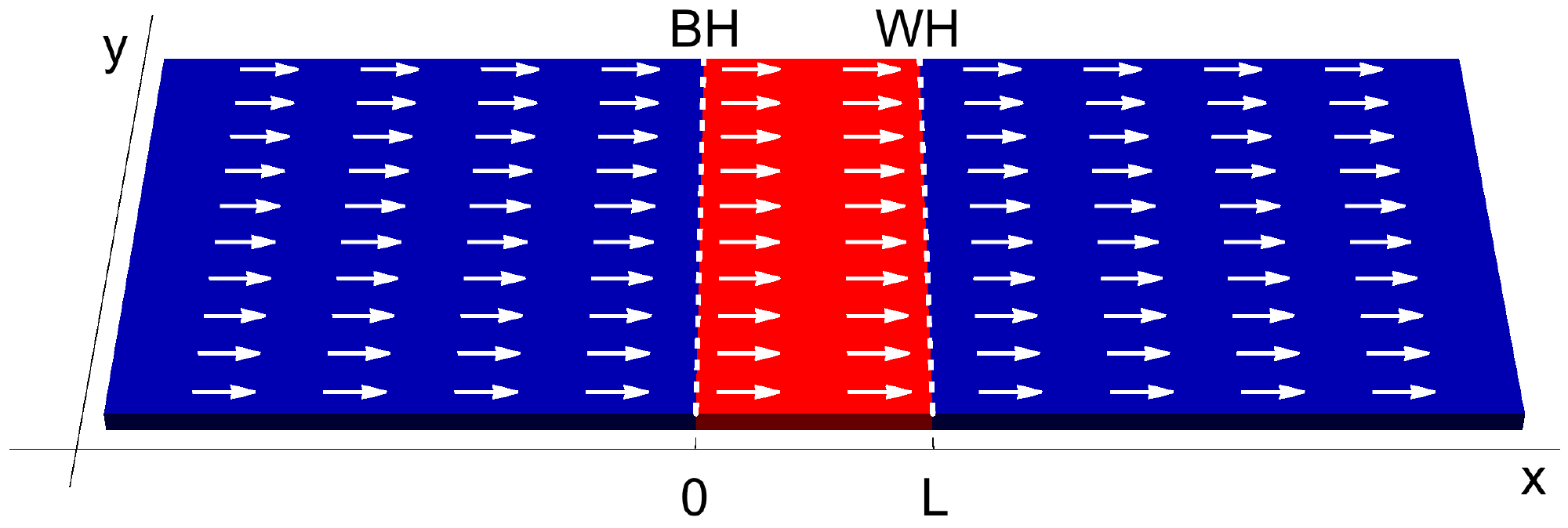}
\caption{Initial setup: Two-dimensional BEC flowing with constant velocity in the $x$-direction. The local speed of sound is tuned in a steplike manner to create a small supersonic (red) region within subsonic flow (blue),  making a black hole horizon at $x=0$ and a white hole horizon at $x=L$.}
\label{fig:setup}
\end{figure}
Following \cite{Garay, Garay2}, we consider a sonic horizon in a dilute BEC at effectively zero temperature. In contrast to previous work on one-dimensional black hole analogs, our system is in (2+1) dimensions. The extra dimension is significant because (1+1)-dimensional black holes lack Bekenstein-Hawking entropy (which is proportional to the horizon measure), and because Unruh's exact mapping between hydrodynamic sound propagation and massless field evolution in curved spacetime is only globally valid in $(D+1)$ dimensions with $D>1$ \cite{secondpaper}. 

In particular we study an idealized scenario in which the initial state of the BEC complex order parameter $\Psi(\bm r,t)$ is a small perturbation around a background $\Psi_{0}$ with uniform density and with uniform velocity $v$ in the $x$-direction. We express all dimensionful quantities in units defined by this initial velocity, so that $t=\hbar \tau / (mv^{2})$ and $\mathbf{r}=(x,y)\hbar/(mv)$, where $m$ is the atomic mass. In these units we write $\Psi_{0}=\alpha \exp(i \bm {v} \bm x) \exp(-i \mu \tau)$ where $\bm {v}=(1,0)$, and $\mu$ and $\alpha$ are constants. Since experimental technology allows the tuning of inter-atomic interactions, the strength of the repulsive contact interaction $g>0$ between the gas particles is assumed to be $x$-dependent, and an external potential $V(x)$ is tuned to compensate for it so that $\mu=1/2 v^2+V(x)+g(x) \alpha^2$ remains $x$-independent. The initial background $\Psi_{0}$ is a stationary solution to the Gross-Pitaevskii equation (GPE)
\begin{equation}
i \frac{\partial \Psi}{\partial \tau}=-\frac 12 \bm {\nabla}^2 \Psi +V(x) \Psi +g(x) |\Psi|^2 \Psi
\end{equation}
which governs the BEC in the mean-field approximation.

$\Psi_{0}$ represents a sonic black-hole/white-hole pair because within the ergoregion $0<x<L$ the lower interaction strength $g(x)$ makes the local speed of sound $c=\sqrt{g\alpha^2}$ lower than the initial background flow velocity $v$. See Fig.~\ref{fig:setup}. As a further idealization, intended to represent a portion of a large horizon, we impose periodic boundary conditions in the $y$-direction; the width of our system is large enough, however, that we detect no finite-size effects. In the $x$-direction our system is effectively infinite: it is periodic but with a length so great that no excitations propagate around it within the duration of our simulation. Since this duration is long enough to see complete relaxation of the initial black hole, the length in $x$ of our system is large enough to be a technical challenge. 

Our final idealization is to make the $x$-dependence of $g$ and $V$ be steplike, so that $g$ takes a constant smaller value inside the ergoregion and a constant larger value outside it. The resulting spatial abruptness of the transitions between sub- and supersonic flow is a simplification which should be removed in future studies, since the Hawking temperature of BEC black holes can only be defined for sufficiently smooth flow velocity and sound speed. One-dimensional studies indicate, however, that the qualitative behavior of BEC sonic horizons is insensitive to the details of the horizon profile. In a smoother background, behavior like that at our abrupt steps in $g$ and $V$ appears instead at semiclassical turning points in perturbations around the background $\Psi_{0}$.
 
\paragraph{\label{bogoliubov} Dynamical Instabilities.---}
The linear stability of a stationary GPE solution like $\Psi_{0}$ is determined by evolving perturbations $\Psi = \Psi_{0}+\delta\Psi$ under Eqn.~(1), while discarding terms of higher than linear order in $\delta{\Psi}$. The thus linearized GPE couples $\delta\Psi$ and $\delta\Psi^{*}$ in a system known as the Bogoliubov-de Gennes equations (BdG). The long wavelength limit of the BdG describes sound waves in a hydrodynamic background and can thus be mapped onto the field equation of a massless field in curved spacetime \cite{Unruh,secondpaper}. At short wavelengths, however, BdG allows propagation faster than the speed of long-wavelength sound. This short wavelength dispersion also implies, moreover, that the eikonal approximation must break down near a sonic horizon, requiring connection formulas which mix short- and long-wavelength modes \cite{corley_jacobson, Garay2}; such mixing also occurs at abrupt horizons like ours \cite{Garay2}.

Analyzing this mixing at a single horizon reveals the remarkable result that, when the perturbations are quantized, the mixing with short-distance modes is not only compatible with long-wavelength Hawking radiation, but is actually the very mechanism by which Hawking radiation occurs in the black hole analog \cite{corley_jacobson}. When a second mixing point is included, however, as is necessary in any finite ergoregion, the doubled mixing leads generically to perturbative modes which grow in time exponentially---dynamical instabilities \cite{corley_jacobson, Garay2}. We emphasize that these are not pathological features of white holes, but are generic for finite ergoregions; they are also present when atom sinks allow purely black holes of finite size \cite{Garay2}. In sonic black holes of $D>1$ spatial dimensions, moreover, the number of dynamically unstable modes tends to be large \cite{secondpaper}.

Since quantum fluctuations forbid growing modes from having zero amplitude, the dynamical instability of a finite laminar black hole means that it cannot actually be a steady state of the system, even with the addition of thermal Hawking radiation. The linear instabilities must grow until they are limited by nonlinearity. Linearized quantum field theory cannot describe this regime, but it can be described at the classical level by the GPE (1).

\paragraph{\label{simulations} Simulations.---}
\begin{figure}
\centering
\includegraphics[width=0.48\textwidth, trim=39mm 9mm 24mm 7mm ,clip]{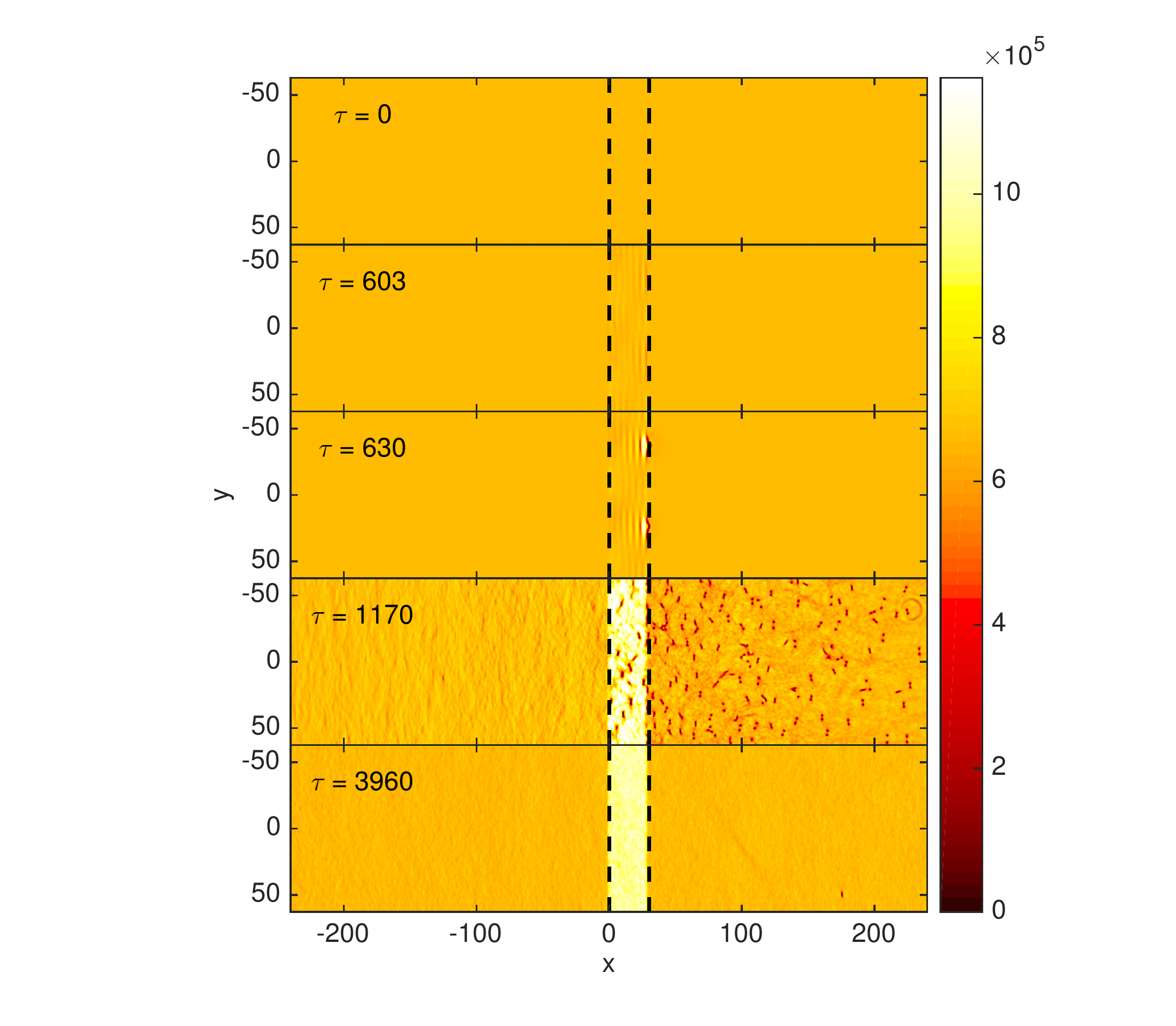}
\caption{Temporal evolution of $|\Psi|^2$ for $c_1=1.5$, $c_2=0.75$, $L=30$ and $\alpha=200/a\sim 815$. The dashed lines indicate the location of the horizons. Only a part of the $x$-range of the system is shown; the total length of the system is $2560\pi$. There is no dissipation; the excitations visible at $\tau=1170$ have simply propagated out of the field of view by $\tau=3960$.}
\label{fig:density}
\end{figure}
\begin{figure*}
\centering
\includegraphics[width=0.90\textwidth, trim=14mm 51mm 17mm 48mm ,clip]{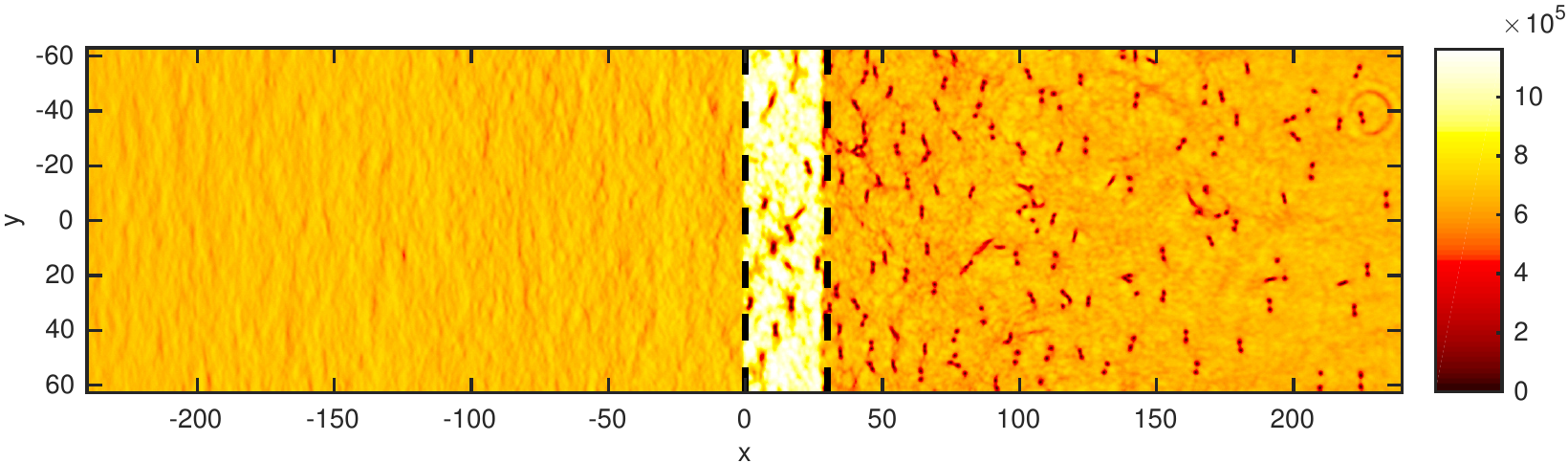}
\caption{Same as Fig.~\ref{fig:density} for $\tau=1170$, enlarged to show detail.}
\label{fig:density_large}
\end{figure*}
\begin{figure*}
\centering
\includegraphics[width=0.90\textwidth, trim=14mm 52mm 16mm 46mm ,clip]{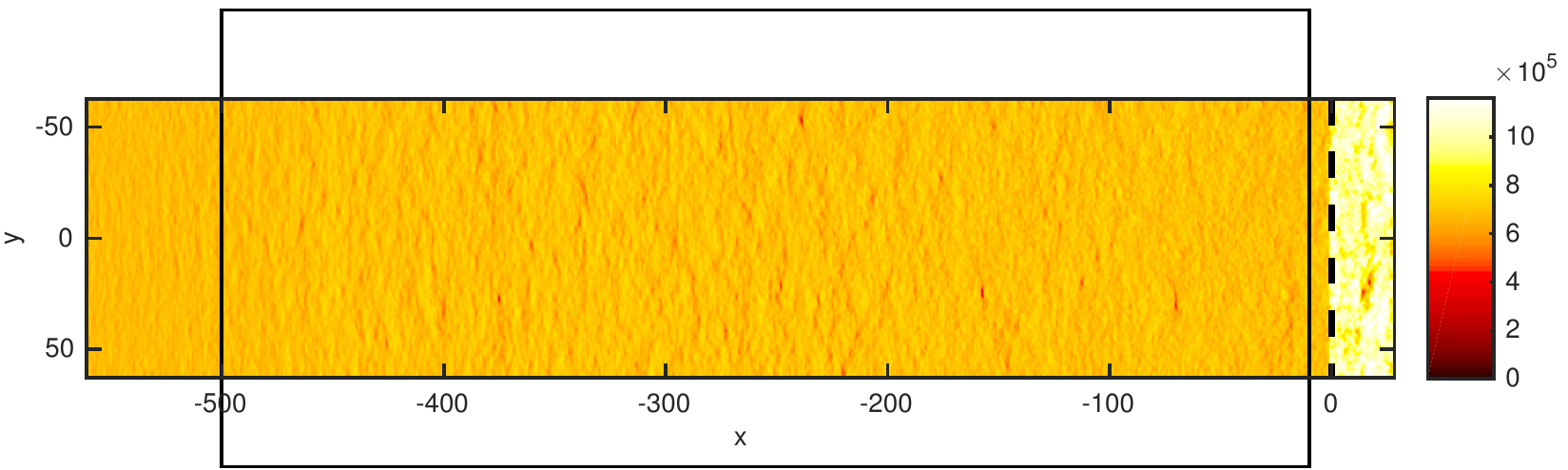}
\caption{Quasi-steady radiation on the black hole side at $\tau=1530$. The box indicates the region that was used to calculate the power spectrum in Figs.~\ref{fig:energydistribution} and \ref{fig:E4}.}
\label{fig:density_box}
\end{figure*}

To investigate the nonlinear regime we use a method similar to the so called Truncated Wigner approximation \cite{sinatra_castin, martin, blakie}: we evolve $\Psi$ classically under (1), but from an initial condition $\Psi(x,y,0)=\sqrt{\rho_{0}+\delta\rho} \exp[i(\theta_0+\delta \theta)]$ with $\delta\rho$, $\delta\theta$ representing quantum fluctuations. These perturbations are random with a probability distribution given by the ground state Wigner function of the BdG excitations in a $\Psi_{0}=\sqrt{\rho_0}\exp(i \theta_0)=\alpha \exp(i\bm v \bm x)$ background with completely uniform $g$ and $V$ equal to the values outside the ergoregion. (See \cite{secondpaper} for a discussion why it is advantageous to switch to the hydrodynamic variables $\rho$ and $\theta$.) This represents an experiment in which a uniform condensate is prepared at zero temperature, and then $g$ and $V$ within $0<x<L$  are suddenly altered to create the ergoregion. 

Since this Gaussian initial state includes fluctuations whose classical energy in each mode below the resolution cut-off would average $\hbar \omega/2$, it provides \cite{secondpaper}
\begin{equation}
\langle \delta \rho^2(\bm x) \rangle=\frac{\rho_0}{L_x L_y}\sum_{\bm k} \frac{|\bm k| \xi}{\sqrt{4+|\bm k|^2 \xi^2}}\simeq \rho_{0}a^{2}
\end{equation}
where $\xi=1/c_1$ is the ambient healing length, $L_{x,y}$ are the system lengths in $x$ and $y$ directions and the sum is over $\bm{k}$ up to the cut-offs represented by the grid spacing $a$. Since the quantum initial state is not stationary under the classical evolution described by the GPE, but thermalizes to a classical distribution with a different temperature \cite{sinatra_castin}, we can only keep the truncated Wigner calculation accurate over our long simulation times by choosing $\rho_{0}$ large enough to keep $\sqrt{\langle\delta\rho^{2}\rangle}/\rho_{0}$ quite small. In the dimensionless units defined above, for which the initial flow speed is 1, our initial state has $|\Psi_{0}|=\alpha=200/a$, corresponding to a very unrealistically high density of approximately $3\times 10^5$ particles per square healing length. This unrealistic requirement of our simulation highlights the difficulty of long-term real-time non-equilibrium nonlinear quantum dynamics: insight from experiment is truly needed. The ergoregion length is $L=30$, $y$-direction period $40\pi$, and $x$-direction period $2560\pi$; $g(x)$ and $V(x)$ are tuned such that the speeds of sound outside and inside the ergoregion are $c_{1}=1.5$ and $c_{2}=0.75$, respectively. These imply that the healing lengths outside and inside are $2/3$ and $4/3$ respectively. Our ground state Wigner function then provides $\sqrt{\langle\delta\rho^{2}\rangle} \sim 0.0048 \rho_0$. Further details of our simulation may be found in \cite{secondpaper}. 

We evolve in time by solving (1) numerically using the standard split-operator fast Fourier transform method \cite{ssfm} on a 32,768$\times$512 grid, so that our grid spacing $a=40 \pi/512$ gives sub-healing-length resolution.  Since long times require short time steps, the numerical task is substantial. We therefore simplify the usual Truncated Wigner method by analyzing only a single realization of the ensemble of initial conditions. Our computation may be taken to represent a single experimental run. As will be seen, the resulting single $\Psi(x,y,\tau)$ appears to realize an ensemble within its evolution, in the sense that spacetime sub-volumes that are not too far apart from each other look like realizations of the same ensemble.
\begin{figure*}
        \centering
        \subfloat{
                \includegraphics[width=0.3\textwidth, trim=0mm 1mm 0mm 0mm ,clip]{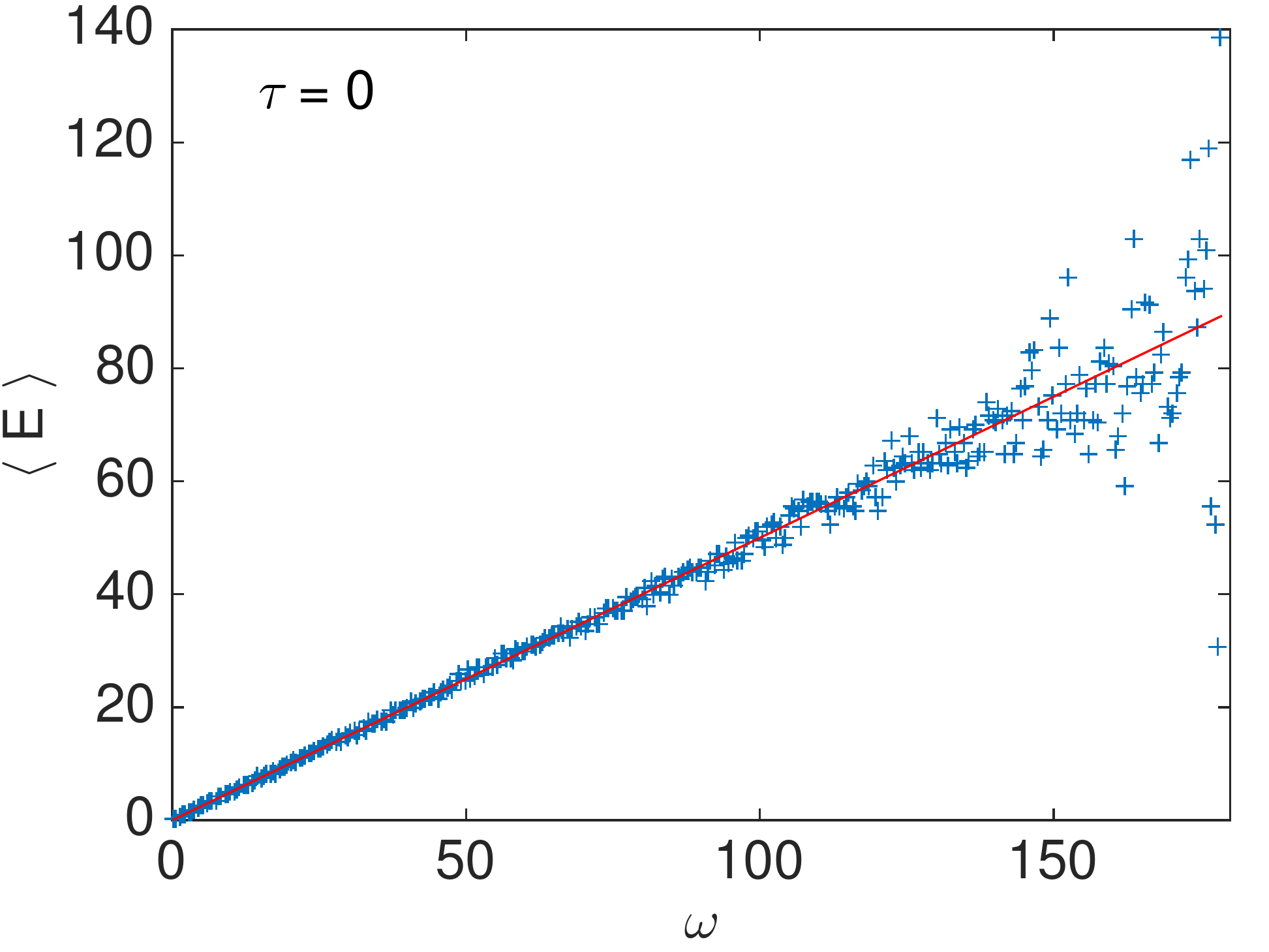}
                \label{fig:E1}
     }
        \subfloat{
                \includegraphics[width=0.3\textwidth, trim=0mm 1mm 0mm 0mm ,clip]{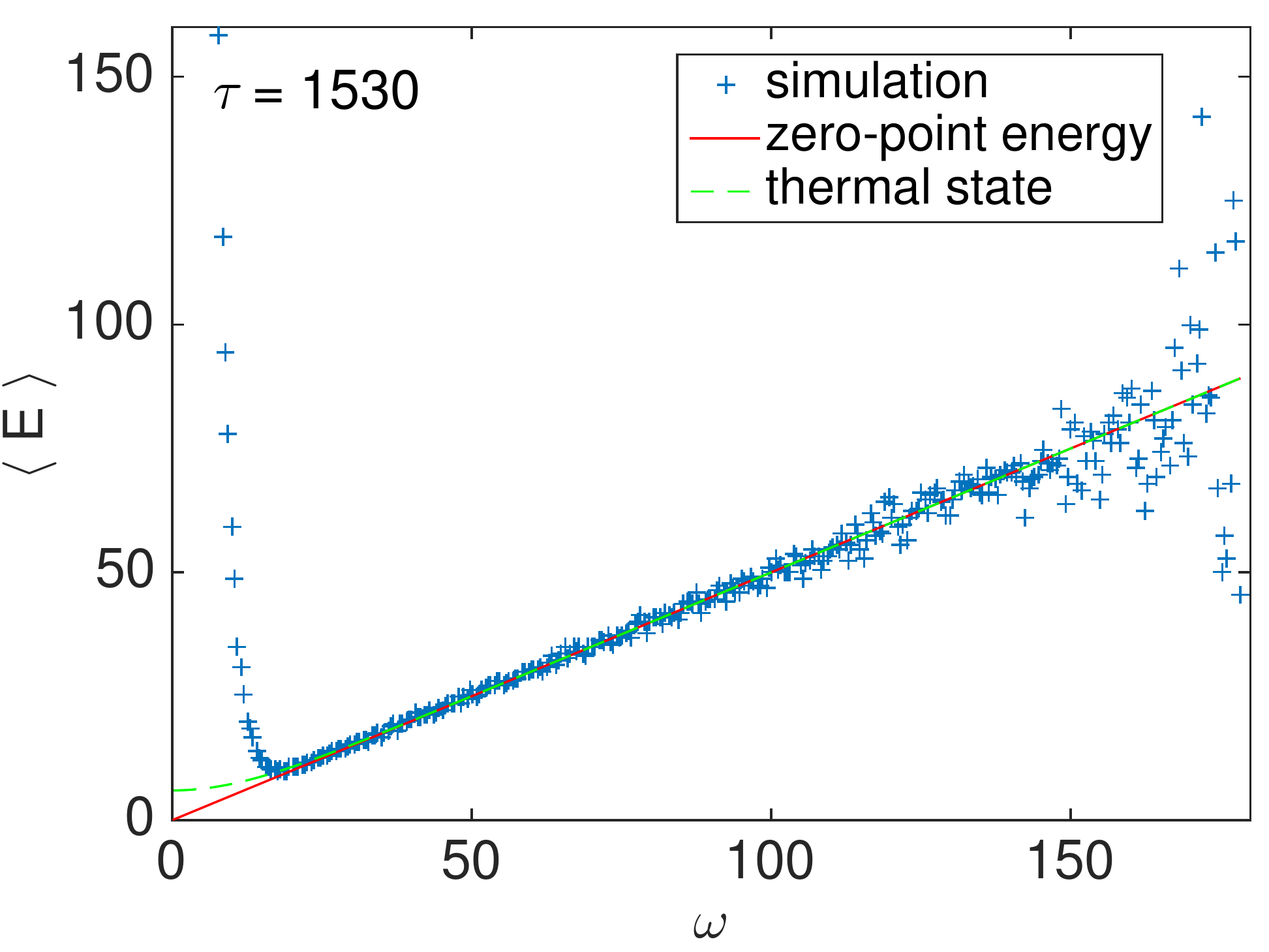}
                \label{fig:E2}
     } 
       \subfloat{
                \includegraphics[width=0.3\textwidth,trim=0mm 0mm 0mm 0mm ,clip]{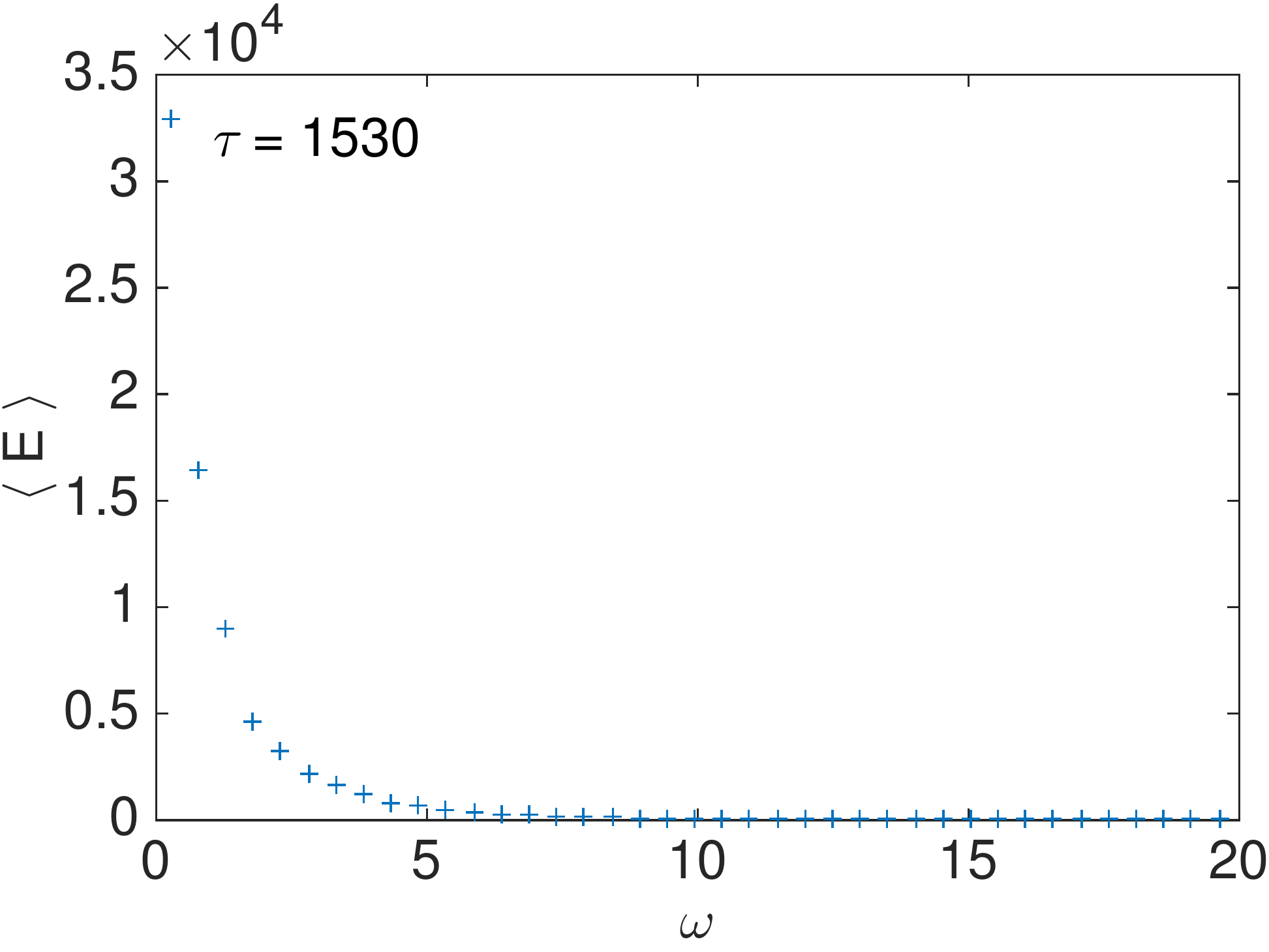} 
                \label{fig:E3}
    }
        \caption{Energy spectrum of the excitations on the black hole side. The solid red line indicates the zero point energy $E=\omega / 2$, the dashed green line indicates the spectrum for a thermal state at $T=6 \,mv^2/k_B$. The increased scatter of the rightmost points is because averaging over frequency bins of constant width gives smaller samples at high frequencies.}
        \label{fig:energydistribution}
\end{figure*}
\begin{figure}
\centering
                \includegraphics[width=0.45\textwidth,trim=0mm 0mm 0mm 0mm ,clip]{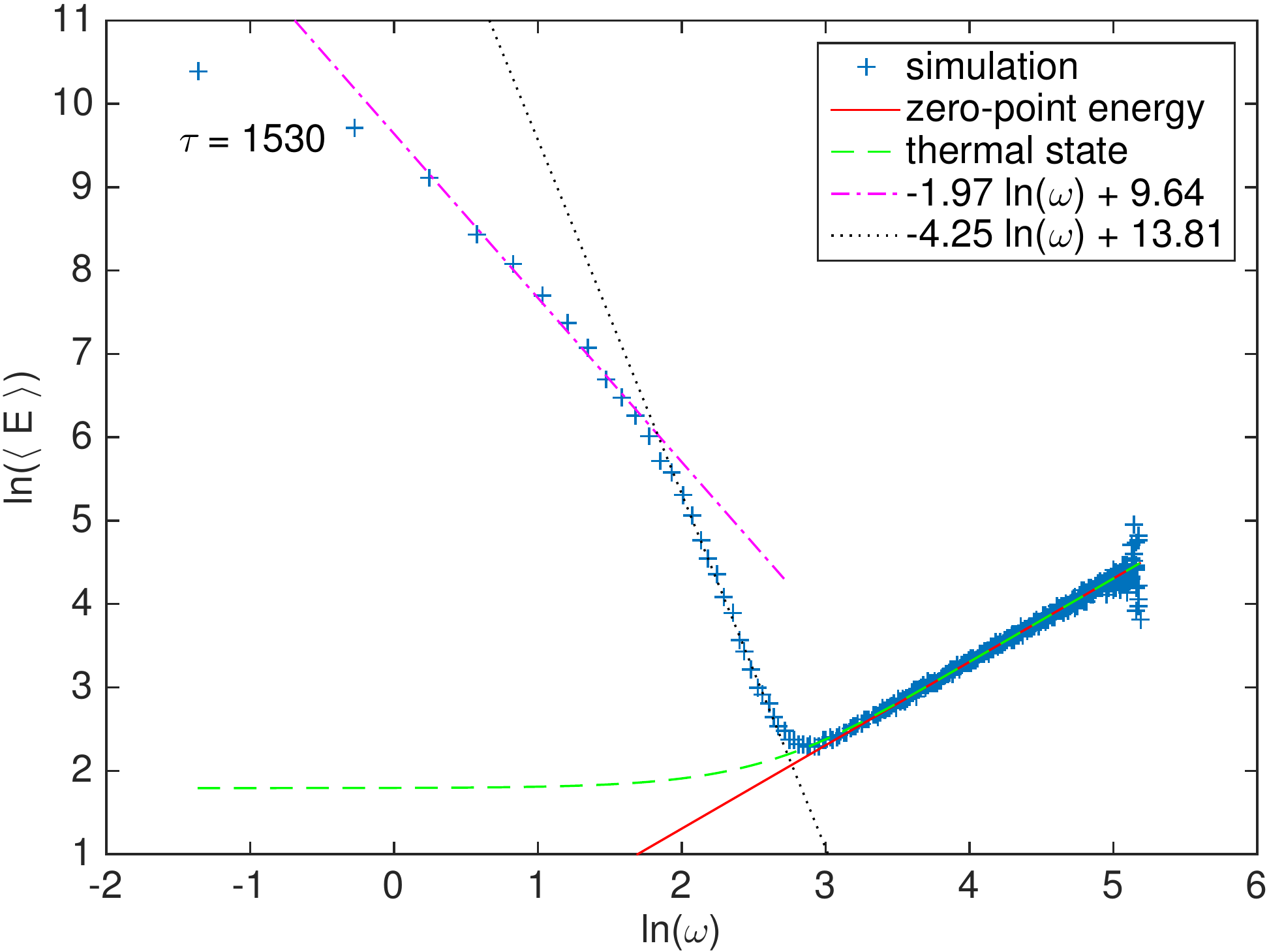}
                \caption{Same as Fig.~\ref{fig:energydistribution} on a double-logarithmic scale. The power spectrum is well described by power laws in three different frequency ranges.}
                 \label{fig:E4}
\end{figure}

The evolution of the density of the condensate can be seen in Fig.~\ref{fig:density}.
In the beginning ($\tau=0$) the perturbations are too small be seen. After some time ($\tau=603$) instabilities seeded by the simulated quantum fluctuations have grown to form a visible pattern of density ripples between the horizons; these then soon deepen into dark solitons which break up into vortex pairs via the so-called snake instability ($\tau=630$). More ripples grow and more vortices form, until the ergoregion becomes a quasi-steady turbulent mess of vortices that continually form and escape (mostly through the white hole horizon). A seemingly random pattern of sound waves is also continuously emitted by this turbulent ergoregion, through both horizons ($\tau=1170$, Fig.~\ref{fig:density_large}). Because the excitations carry away energy from the supersonic region, the flow there gradually slows, while the condensate density in the ergoregion slowly rises to maintain continuity. Eventually the background velocity in the ergoregion falls below the local speed of sound, which rises with background density; the sonic horizons disappear, the subsonic superfluid flow is stable, and vortex production ceases ($\tau=3960$). (The emitted vortices and sound waves continue to move away along our long $x$-axis, outside the Fig.~\ref{fig:density} field of view.)

This eventual decay of the sonic ergoregion resembles the expected Hawking evaporation of an isolated black hole. Our attention focuses, however, on the long quasi-steady phase of emission from the turbulent ergoregion, as shown enlarged in Fig.~\ref{fig:density_large}. Since it is the sound emitted from the black hole horizon which is the analog of Hawking radiation, we analyze the radiation field on the black horizon side in more detail, leaving study of the vortex-rich regions, in which perturbations are too large for BdG theory, for future work. 

\paragraph{\label{steady} Steady State.---}
To analyze the steady emission quantitatively we choose a time ($\tau=1530$) at which the emitted wave pattern appears homogeneously random within a large region, shown in Fig.~\ref{fig:density_box}. We then decompose $\delta \rho$ and $\delta \theta$ in Fourier modes, and count their energies, binned in $k$-space, as functions of BdG frequency. The results are shown Fig.~\ref{fig:energydistribution}, with the analogous data from $\tau=0$ for comparison. At $\tau=0$ the energies follow the line $\langle E\rangle = \omega/2$ because the Wigner function from which they were randomly drawn provides this zero-point energy; the data points scatter more at higher frequencies because there are fewer points in each $\omega$-bin ($\omega\propto k^{2}$). The high-frequency energy distribution remains essentially unchanged in the steady-state epoch at $\tau=1530$, but so much power has been generated at long wavelengths that we need two separate scales to see all of it. A log-log plot (Fig.~\ref{fig:E4}) seems to reveal two distinct power laws in addition to the zero-point energy at high frequencies. The points where the power laws turn over into each other fix two characteristic frequencies $\omega_1\sim6.2$ and $\omega_2\sim15.8$, which correspond to wavelengths on the order of (or somewhat shorter than) the healing length. Sound emitted by turbulent regions is a subject in aeronautical engineering \cite{aero_Lighthill1, aero_Lighthill2, aero_Dowling}, and superfluid turbulence is an active research topic in physics \cite{bagnato}, but the sound emitted by a turbulent superfluid ergoregion does not yet seem to have a theory with which we can compare our results. 

The sonic radiation whose power spectrum is shown in Figs.~\ref{fig:energydistribution} and \ref{fig:E4} can be called Hawking radiation in the loose sense that it is radiation emitted from a horizon. The long-wavelength radiation emitted by the sonic black hole horizon is clearly not thermal, however (at least not for a global temperature---see \cite{secondpaper} for a separate analysis of small regions in $k$-space). Our results therefore show that the turbulent instabilities that were already anticipated in Unruh's first paper \cite{Unruh} cannot be neglected as a mere technical nuisance, for they determine the steady state of the system even at long wavelengths. Indeed it may be more accurate to say that noise emitted from the turbulent ergoregion is what sonic black holes have instead of thermal Hawking radiation. If similar phenomena arise in the more strongly quantum mechanical evolution of real condensates at much lower density than our model, the turbulent ergoregion and its loud zero sound noise should be observable in experiments more easily than the faint thermal Hawking radiation that is predicted when dynamical instabilities are absent.

\paragraph{\label{conclusion} Horizon Physics.---}
Superfluid turbulence in BECs involves short-wavelength nonlinear dynamics beyond the long-wavelength linear analogy with quantum fields in curved spacetime. There is no reason to expect real black holes to resemble turbulent ergoregions. Why should we then care about sonic black holes? We answer that real black holes are only one particular case of quasi-steady event horizons in a long-wavelength theory. To learn that they are typical of all such horizons would be to learn something significant about black holes---as it would be to learn instead that they are a special case.

And in any case the greatest significance of Hawking radiation is not in adding one more weird feature to distant compact objects, but in suggesting a non-statistical connection between macroscopic steady states (\textit{i.e.} thermodynamics) and pure geometry (horizons). Analog black holes may shed light on this connection, even if they do not otherwise resemble real black holes. For instance, the roughly uniform density of vortices within our ergoregion, even as they steadily drift through the white horizon, seems to indicate steady production of vortices along the black horizon, which might suggest an analog of Bekenstein-Hawking entropy. 

\begin{acknowledgments}
The authors acknowledge support from the Landesforschungszentrum OPTIMAS and the Deutsche Forschungsgemeinschaft (DFG) through the SFB/TR185 (OSCAR).
\end{acknowledgments}

\bibliography{citations}

\end{document}